\documentclass{aa}  

\usepackage{graphicx}
\usepackage{txfonts}
\usepackage{breqn}
\usepackage{hyperref}
\begin{document} 

\title{Interstellar Meteors from Tidal Disruption of Rocky Planets on Eccentric Orbits Around M-Dwarfs}

\titlerunning{Tidal Disruption of Rocky Planets}

\author{Abraham Loeb\inst{1} and Morgan MacLeod\inst{1}}
\institute{\inst{1}Institute for Theory and Computation, Center for Astrophysics, Harvard \& Smithsonian, 60 Garden Street, Cambridge, MA, 02138}

\date{submitted: \today}

% \abstract{}{}{}{}{} 
% 5 {} token are mandatory
 
  \abstract
   {Low-mass stars appear to frequently host planetary systems. When these rocky planets develop high eccentricities as a result of secular torques or dynamical scatterings, they occasionally pass close to the host star. In these close passages, planets can be tidally disrupted, and sheared into bound and unbound debris tails. To suffer such a disruption the stellar density must by higher than the planetary density.} 
   {This condition is met for the most common star and planet systems, M-dwarf stars hosting rocky planets. We describe the dynamics of a tidal disruption, and estimate the typical velocities of unbound ejecta.}
   {We simulate the gas dynamics of a planetary tidal disruption, and show that disruptions preserve the layered structure of a rocky body, with the outermost layers flung into interstellar space with the highest velocities.} 
   {We compare these properties to that of the candidate interstellar meteoroid CNEOS-2014-01-08 (IM1). IM1's approximately 60~km~s$^{-1}$  excess speed relative to the local standard of rest is naturally reproduced by the unbound debris of the disruption of an Earth-like planet around an M-dwarf star. We suggest that such an encounter might explain the interstellar kinematics of IM1, and its unusual composition, especially if it originated in the fastest-expelled crust of a differentiated rocky planet. Finally, we estimate that the disruption of $\sim 10M_\oplus$ reservoirs of rocky planets per M-dwarf are needed to reproduce the inferred rate of IM1-like objects. }
 {}


   \maketitle
%
%________________________________________________________________
\section{Introduction} \label{sec:intro}

Stars or planets can be disrupted by tides when they pass close enough to another massive object that the tidal force exceeds their own self-binding force. This happens to stars around black holes in galactic centers \citep[e.g.][]{1988Natur.333..523R}, and can also happen when rocky bodies like planets or asteroids pass near their host stars \citep[e.g.][]{2018ApJ...861...35R}. 

Evidence for tidal disruption of rocky bodies near dwarf stars is
most apparent in studies of white-dwarfs (WDs). 
Observations indicate that $\sim 25\%$-–$50\%$ of all
WDs exhibit spectral lines that are indicative of the presence of
metals in their atmospheres~\citep{Zuc03,Zuc10,Ko14,Bro23}.
The metal enrichment is consistent with the
composition of rocky material~\citep{Zuc07,Ga12,Far13,Jura14},
suggesting that WD pollution originates
from minor rocky bodies, with unbound material 
enriching interstellar space~\citep{Per2,Per1}. 
The observed fraction of
polluted WDs and the level of pollution was claimed not to change with
the WD cooling age~\citep{Ko14,Wy14}. Conflicting suggestions~\citep{Hol18} were refuted recently~\citep{Blo22}.  

One would expect rocky planets and asteroids that get within approximately a solar radius distance from  dwarf stars to get tidally disrupted and accreted. 
This expectation is supported by observations of circumstellar disks,
revealed by infrared excess in the stellar spectrum around
polluted WDs~\citep{Far16}.  The formation timescale of a
debris disk following disruption is short compared to the cooling
ages of most polluted WDs~\citep{Veras14,Veras15}, and
all WDs with detected disks have atmospheric
pollution. The association with tidal disruption is directly implied by the
observation of minor bodies transiting the polluted WD
1145+017~\citep{Van15a,Alonso16,Gan16,Rap16,Xu16}  as well as in 
ZTF J0139+5245, ZTF J0328-1219 and WD 1054-226 
(see Figures 2 \& 6
in~\citet{Ver21} and the penultimate paragraph of Section 4.2 in~\citet{Ver24}). 

The exact physical processes that fuel the disruption of rocky bodies around WDs remains unknown, though many plausible scenarios have been suggested. Planetary dynamical instabilities can lead to the observed
tidal disruption of rocky bodies by dwarf stars~\citep{Deb02,Veras13,Mus14,Ver15}.
Stellar mass loss can also widen the region
around mean-motion resonances where chaotic diffusion of
rocky planets and asteroids is efficient, leading to tidal
disruption~\citep{Bon11,Deb12,Fre14}. This may be particularly relevant to disruptions of bodies around WDs, which have lost considerable mass in their formation. The likelihood of disruption increases in the presence of binary
companions~\citep{Kra12}, which are known to be abundant even for stellar remnants like WDs~\citep{Shahaf2023}.

Earth-mass planets 
could develop high eccentricity orbits as a result of a secular torque from an outer giant planet~\citep{Morales19} or a stellar binary companion~\citep{PM17,Van2020}. The resulting secular instabilities or the 
Kozai-Lidov mechanism~\citep{Koz62,Lid62,Naoz2016}, could
trigger tidal disruptions of bodies that feed the M-dwarf with rocky
material.

%Here, we adopt the association of the ``BeLaU"-type abundance pattern with a highly differentiated magma ocean of a rocky planet with an iron core that is tidally heated during many tight periapse passages before being disrupted by an M-dwarf, the most common type of stars with about a tenth of the mass of the Sun.~\footnote{We note that \citet{2018ApJ...861...35R} and \citet{Zhang20} had considered a tidal disruption scenario as a possible formation mechanism for 1I/2017 U1 'Oumuamua.} 

%\section{Pollution of Dwarf Stars}\label{sec:pollution}

Here, we describe how the tidal disruption of planets and rocky bodies that are in highly eccentric orbits generates bound and unbound tails of debris. The bound tail may eventually accrete onto the host star, while the unbound tail escapes the gravity of the host star. We consider how this process might contribute to a population of unbound, interstellar rocky objects, which might pass through the solar system or collide with Earth.

\section{Tidal Disruption of Rocky Planets by M-Dwarfs}\label{sec:TD}

Planets can be disrupted by tides around their host dwarf stars only when they pass extremely close at periapse. Since these orbits are destructive to the planet, dynamical processes are needed to feed planets or other rocky bodies into highly eccentric orbits that lead to their disruption.

\subsection{Disruption Criteria}

For a stellar mass $M_\star$ and a planetary mass $m_{\rm planet}$, tidal disruption 
occurs at a periapse distance of the ``tidal radius", 
\begin{equation}
r_{\rm t} = \left({M_\star \over m_{\rm planet}}\right)^{1/3}R_{\rm planet} = \left({M_\star \over {4\over 3}\pi \rho_{\rm planet}}\right)^{1/3} 
\label{eq:tide}
\end{equation}
which gives
\begin{equation}
r_{\rm t} \approx 0.6R_\odot\left({M_\star\over0.5M_\odot}\right)^{1/3} \left({\rho_{\rm planet}\over  3~{\rm g~cm^{-3}} }\right)^{-1/3} ,
\label{eq:tidescale}
\end{equation}
where we have substituted a mean density of $\rho_{\rm planet}=(3m_{\rm planet}/4\pi R_{\rm planet}^3)=3~{\rm g~cm^{-3}}$ and $R_{\rm planet}$ is the planetary radius. By comparison, Earth's mean density is $\sim 5.5~{\rm g~cm^{-3}}$ while the Moon is  $\sim 3.3~{\rm g~cm^{-3}}$ and Ceres is $\sim 2.1~{\rm g~cm^{-3}}$. 

The observed radius of M-dwarfs scales linearly with mass, $R_\star\sim 0.1 R_\odot (M_\star/0.1M_\odot)$~\citep{Parsons18}, implying that the tidal disruption radius is larger than the stellar radius by a factor of $\sim 3 ({M_\star/0.1M_\odot})^{-2/3}$. Thus, M-dwarfs can disrupt rocks that get within a few times their photospheric radius without swallowing them. Swallowing occurs because of gaseous drag when planets pass significantly within the stellar photosphere \citep[e.g.][]{2018ApJ...853L...1M}.

\begin{figure}
    \centering
    \includegraphics[width=0.45\textwidth]{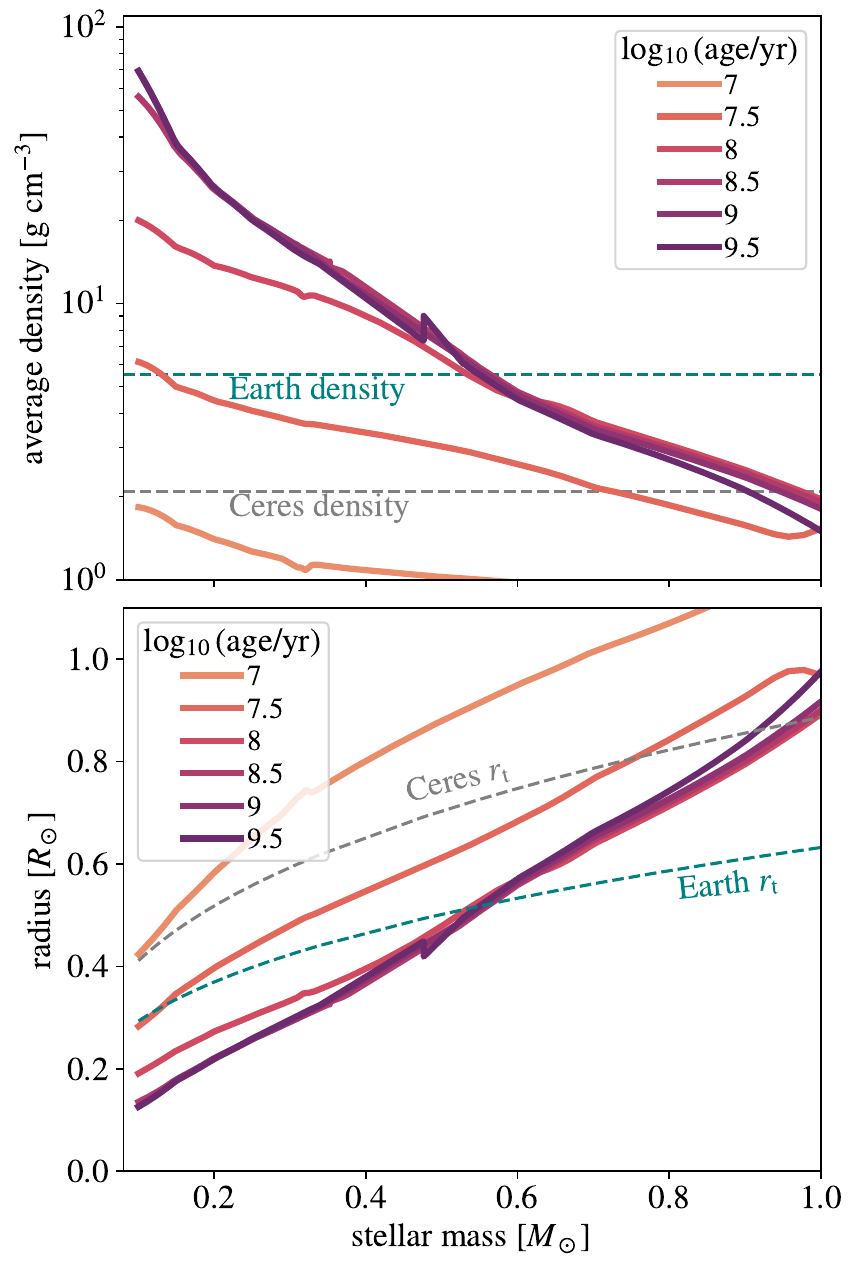}
    \caption{Model isochrones from MIST \citep{2016ApJ...823..102C} of varying ages, shown in terms of their average density (upper panel) and radius (lower panel). The upper panel compares stellar mean densities to those of Ceres and Earth. Stars denser than typical rocky bodies can disrupt these objects by tides. The lower panel compares the tidal radius,  $r_{\rm t}$ (equation \ref{eq:tide}), to stellar radii in units of $R_\odot$. In this case, stars more compact than $r_{\rm t}$ can disrupt an object without swallowing it whole. These isochrones show that as stars cool and collapse to their main sequence radii, low-mass stars ($\lesssim 0.5M_\odot$) become denser than typical rocky planets and planetesimals, meaning that they can tidally disrupt these objects. }
    \label{fig:isochrone}
\end{figure}

Figure \ref{fig:isochrone} shows Mesa Isochrones and Stellar Tracks (MIST) \citep{2016ApJ...823..102C} model isochrones for low-mass stars on the pre-main sequence and main sequence. As these stars cool and collapse, they become denser, such that their mean densities exceed that of typical rocky bodies. This means that these low-mass stars (less than approximately $0.5M_\odot$) can disrupt rocky planets and planetesimals without swallowing them whole. By contrast, higher mass stars have lower mean densities, as do very young pre-main sequence stars. Thus, even $0.1M_\odot$ stars, if they are young enough, with ages $\lesssim10^{7.5}$~yr, would swallow rather than disrupt their planets.

\subsection{Debris of Tidal Disruption}

The debris of tidal disruption is imparted with a spread in orbital binding energies relative to the disrupting object \citep[e.g.][]{1988Natur.333..523R}. To first order, this spread in energy per unit mass at periapse is 
\begin{equation}
\Delta E \sim \left({G M_\star \over r_{\rm peri}}\right) \times \left( {R_{\rm planet} \over r_{\rm peri}}\right), 
\label{eq:E}
\end{equation}
for a periapse distance $r_{\rm peri} \sim r_{\rm t}$. 
If we imagine an initially parabolic encounter between the star and planet (i.e., the orbital energy is zero), then some material ends up bound to the star, with a binding energy of up to $-\Delta E$ and some ends up unbound, with a excess energy of up to $+\Delta E$. This spread in energy, imparted at periapse, leads tidal debris to spread in time, forming a long, thin stream of bound and unbound material. 

Bound material streams back to the vicinity of the periapse passage, possibly self-intersecting, colliding and forming a debris disk around the host star. This debris disk might be expected to have a fragment size distribution similar to the bound debris observed around WDs~\citep{Van15a,Ver21,Ver24}.  However, accreted material onto an M-dwarf would form a small fraction of the metal content of the star, making the ``pollution" which is evident on WDs (see Introduction) nearly invisible in the context of an M-dwarf accretor. In particular, planetary material would be mixed through the convective region of the star \citep[e.g.][]{2021AJ....162..273S}. In a fully-convective M-dwarf, the added metals would form a fractional contribution of only $\sim m_{\rm planet} / (Z M_*) \sim 10^{-3}$ where $Z$ is the metal fraction and the numerical value comes from adopting $m_{\rm planet} \sim M_\oplus$, $Z=0.02$, and $M_* =0.1M_\odot$. 

Because planet or planetesimal orbits are typically bound to the host star, unbound debris are generated only when the spread in energy imparted at peripase passage exceeds the original orbital energy of the rocky planet, 
\begin{equation}
E_{\rm orb} =  - {G M_\star \over 2 a}, 
\label{eq:DE}
\end{equation}
where $a$ is the orbital semi-major axis. Thus, $E_{\rm orb} + \Delta E > 0$ (or equivalently $\Delta E > |E_{\rm orb}|$)  is the critical condition for unbound debris. 
We therefore find a condition on the semi-major axis of 
\begin{equation}
a > { r_{\rm peri}^2 \over 2 R_{\rm planet} } \sim {1 \over 2} \left({M_\star \over m_{\rm planet}}\right)^{2/3}R_{\rm planet}
\end{equation}
for the generation of unbound debris in a tidal encounter, where the second equality assumes $r_{\rm peri} \sim r_{\rm t}$. Otherwise all of the tidal debris is bound to the host star \citep[e.g.][]{2013MNRAS.434..909H,2018ApJ...861...35R}.  For an Earth-like planet around a $0.1M_\odot$ star, we find that orbits with $a \gtrsim 0.04$~au (or orbital periods $P_{\rm orb} \gtrsim 10$~d) yield unbound debris. The implied eccentricity is 
\begin{equation}
    e \gtrsim 1-\left({m_{\rm planet}\over M_\star}\right)^{1/3},
\end{equation}
or $e\gtrsim0.97$ for an Earth-like planet. 

This implies that planets on typical orbits, when deviated to high eccentricity and disrupted by tides generate unbound debris, which is free to propagate into interstellar space. The asymptotic velocity of the debris is given by 
\begin{equation}\label{eq:vinf}
    v_{\infty} \approx \sqrt{2 (E_{\rm orb} + \Delta E) }
\end{equation}
In the limit that $\Delta E \gg |E_{\rm orb}|$, this simplifies to $v_{\infty} \sim \sqrt{2  \Delta E }$, or in terms of typical quantities, 
\begin{dmath}
    v_\infty \sim 60 {\rm \  km \ s}^{-1} \left( M_* \over 0.1 M_\odot \right)^{1/6} \left( m_{\rm planet} \over  M_\oplus \right)^{1/3}  \left( R_{\rm planet} \over  R_\oplus \right)^{-1/2} \left( r_{\rm peri} \over  r_{\rm t} \right)^{-1} .
\label{eq:60}
\end{dmath}

\section{Simulated Planetary Disruption}\label{sec:sim}

To examine how a tidal disruption leads to unbound debris and where different asymptotic velocities originate in the planet debris, we followed a calculation of a model planet disrupted by its host star. 

\subsection{Method}

We simulate the tidal disruption of a planet by its host star, treating the planet as a fluid in the Athena++ hydrodynamic code \citep{2020ApJS..249....4S}. We solve the equations of inviscid hydrodynamics in the reference frame of the planet with additional source terms for the non-inertial reference frame, gravity of the planetary core, and stellar gravity. The planetary material's self-gravity is treated in the monopole approximation \citep{2022ApJ...937...37M}, which may lead the debris streams to be spatially broader than might be expected were they self-gravitating, but is not thought to otherwise effect the results of our study. The setup is publicly available and identical to that utilized by \citet{2022ApJ...937...37M} and \citet{2023NatAs.tmp..183M}. 

To model the planet, we adopt a relatively incompressible fluid structure. We adopt polytropic index $\Gamma=5$, where $P\propto \rho^{\Gamma}$ within the polytrope structure. The inner 50\% of the model is excised to represent the iron core, while the outer 50\% is treated as a fluid with an ideal gas equation of state with an adiabatic index $\gamma=5$. The initial model roughly matches the density profile of Earth's mantle. This simple equation of state does not capture the material strength, phase transitions, or other relevant physics of rocky matter at planetary temperatures and therefore is not strictly realistic. However, studies of tidal disruption dynamics find that the tidal gravitational field is a dominant effect in the disruption kinematics \citep[e.g.][]{2013MNRAS.435.1809S}, with the compressibility of the fluid being a secondary factor \citep[e.g.][]{2013ApJ...767...25G}. We use this approximation to capture the approximate behavior through periapse of a relatively incompressible liquid or solid planet. 

Our model system adopts a stellar mass, of $0.1 M_\odot$, a planetary mass of $1M_\oplus$, a planetary radius of $1R_\oplus$, a semi-major axis of $3\times 10^{12}$~cm~$=0.2$~au, and an eccentricity of $e=0.99$. The periapse distance in this configuration, $r_{\rm peri}=3\times 10^{10}$~cm is approximately $1.45 r_{\rm t}$. Our calculation is performed on a static spherical-polar mesh surrounding the planetary core. It extends across $0.5R_\oplus < r < 100R_\oplus$ in the radial direction, covered by 192 logarithmically-spaced zones. The angular directions span the full $4\pi$ of solid angle, from $0 < \theta < \pi$ and $0<\phi < 2\pi$ with 96 and 192 zones respectively. 

\subsection{Results}

\begin{figure*}
    \centering
    \includegraphics[width=\textwidth]{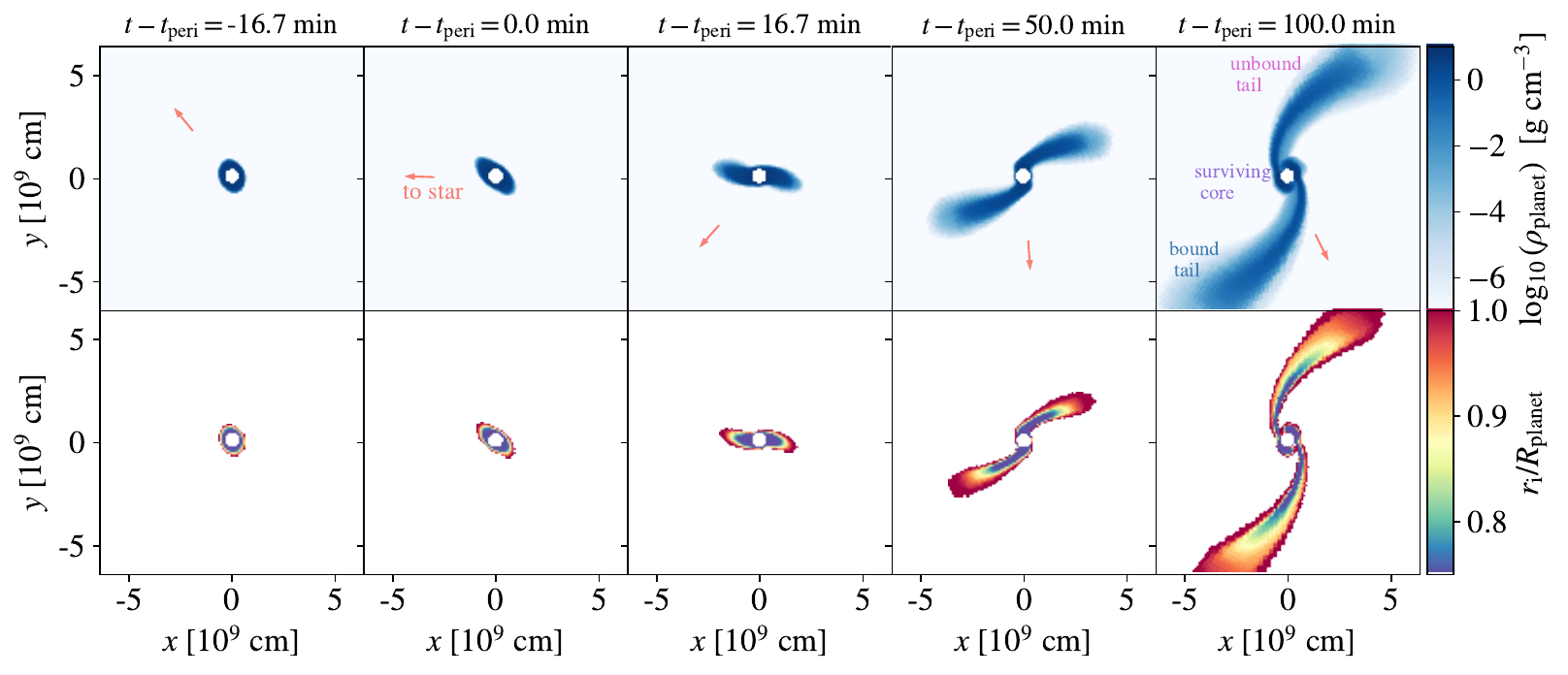}
    \caption{Time series of the tidal disruption of an approximately Earth-like planet as it flies too close to its $0.1M_\odot$ host star. Times are annotated relative to the time of periapse passage, $t_{\rm peri}$. At closest approach, the planet passes at $\sim 1.45r_{\rm t}$, close enough to disrupt the outer layers while the inner core survives the encounter. The iron core is excised in our calculation, and the infall of self-bound material back to the surviving core is visible in the last panel. While the upper series of panels plots the volume-averaged density of planetary debris (in which rocks could maintain a high internal density), the lower shows material's initial radius $r_i$ within the planet relative to the planetary radius, $R_{\rm planet}$. Here, for example, the outermost layers from the planet's crust (shown in red) form the outermost debris in the tidal tails.  }
    \label{fig:peri}
\end{figure*}

Figure \ref{fig:peri} follows the planet through its partially disruptive periapse passage. These frames are slices through the orbital plane and they stay centered on the planet even as it flies past the star. Arrows indicate the direction of the star, which passes at a minimum distance of $3\times 10^{10}$~cm, well outside the frame of view. 
The upper panel shows the volume-averaged density of planetary debris (which could be populated by individual rocks), $\rho_{\rm planet}$, while the lower panel shows the initial radius, $r_i$, of material within the planet relative to the planetary radius. 

Through the approximately hour-long periapse passage, the planet is stretched and distorted. The outer layers are stripped from the planet into two symmetric tidal tails. A hundred minutes after periapse passage, three distinct structures have formed. Those these timescales of periapse passage are unique to this particular scenario, they scale with the orbital timescale at periapse, approximately $(r_{\rm peri}^3 / G M*)^{1/2}$.  Two discrete tidal tails stretch and spread primarily under the influence of the stellar gravity -- expansion having rendered their own pressure relatively unimportant \citep{1988Natur.333..523R}. Over time, they may be confined to narrow, columnar streams by self-gravity \citep{1994ApJ...422..508K,2022ApJ...931L...6B,2023MNRAS.522.5500C,2023MNRAS.526.2323F}. However, their distribution of binding energy relative to the star is already imprinted at this stage \citep{2005Icar..175..248F,2011ApJ...732...74G,2013MNRAS.435.1809S}. 
One tail, that facing away from the star at 100 minutes post-periapse, is unbound, and will fly into interstellar space. Another tail is bound to the star. It will fall back to the vicinity of its original periapse distance at a few stellar radii, where it likely self-intersects and forms a debris disk surrounding the star. 

The slices tracing the original radius of material in Figure \ref{fig:peri} are computed using a passive scalar tracer fluid during the calculation. We see that the stratification of the planet is largely preserved in the tidal tails. The outermost material from the planet crust is flung furthest into the tails. In the particular case we simulate, the unbound and bound debris tails are formed from material at $r_i \gtrsim 0.8 R_{\rm planet}$, perhaps representing the outer mantle and crust of an Earth-mass planet. The surviving core is primarily material from deeper in the planetary interior, including the core (explicitly excised in our calculation) and inner mantle layers from 50-80\% of $R_{\rm planet}$. In the surviving core, some mixing does occur as material initially in the streams but self-bound to the core falls back, a process beginning at 100 minutes post-periapse.

\begin{figure*}
    \centering
    \includegraphics[width=\textwidth]{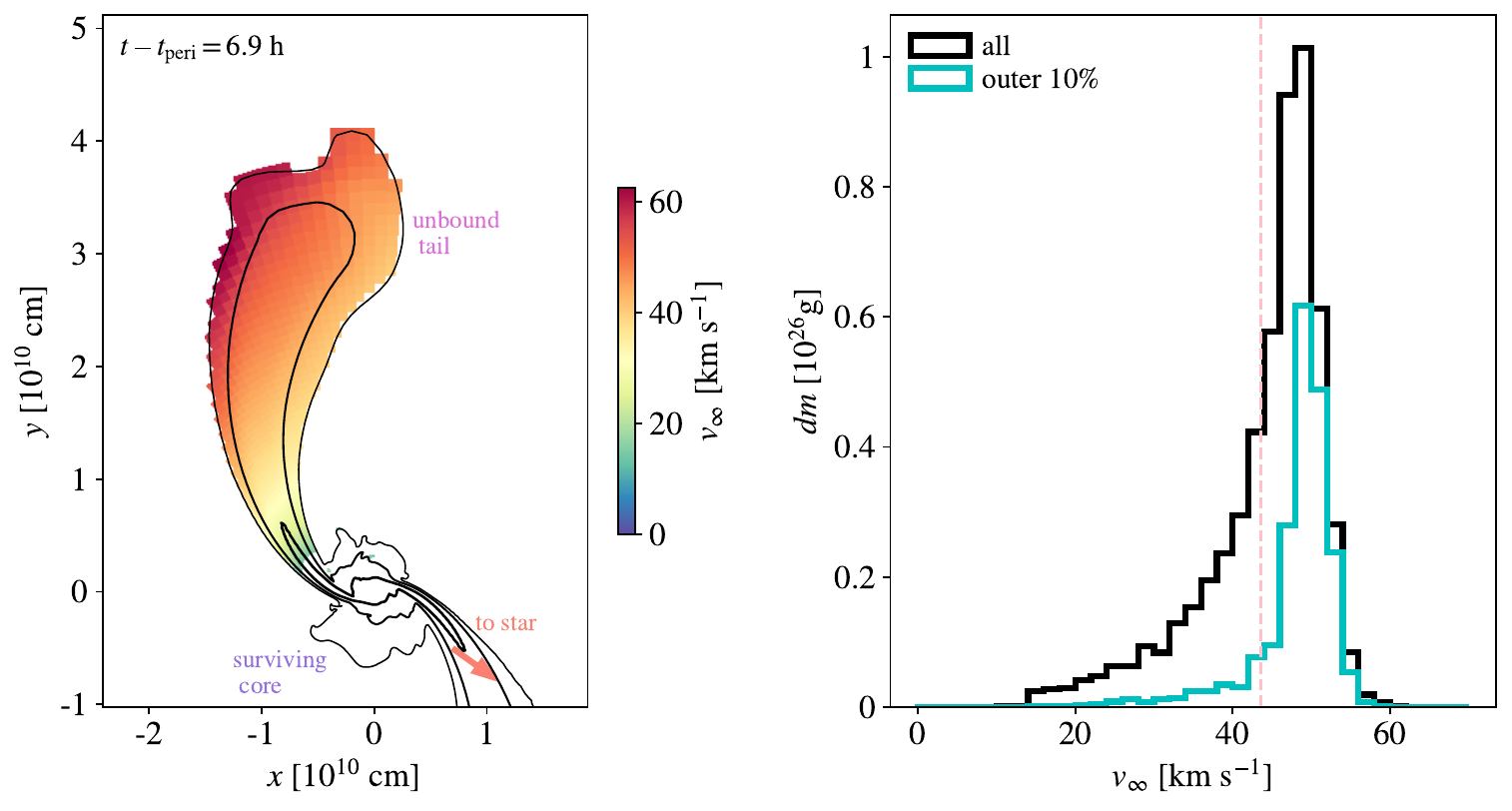}
    \caption{Examination of the asymptotic velocity distribution of the unbound tail. The left-hand panel highlights the spatial distribution of velocity within unbound material, overlaid with contours of the density at $\rho_{\rm planet} = 10^{-6}$, $10^{-4}$, and $10^{-2}$~g~cm$^{-3}$. The velocity histograms in the right panel are compared to the analytic estimate of $v_\infty \sim \sqrt{2\Delta E}$, shown with a vertical dashed line. A comparison between all of the unbound debris and that originating in the outer 10\% of the original planetary radius shows that the outermost layers form the fastest moving ejecta. The interface between composition and kinematics are important in chemically differentiated planets because it determines which layers are expelled with a certain velocity.  }
    \label{fig:vinf}
\end{figure*}

In Figure \ref{fig:vinf}, we turn our attention to the asymptotic velocity distribution imprinted on the unbound tidal debris tail. Contours in Figure \ref{fig:vinf} show density within the tidal tail approximately $\sim 7$~h after periapse passage. Within the unbound tail, we report the asymptotic velocity of material based on its excess kinetic energy in equation (\ref{eq:vinf}). While the left panel shows a slice in the orbital plane, the right panel shows a histogram including the full volume of the unbound tail. 

Figure \ref{fig:vinf} shows that the tidal debris tail represents a broad distribution of velocities. Already in the initial passage (e.g., at 100 minutes post-periapse in Figure \ref{fig:peri}) the fastest-moving material is furthest flung into the tails. We see that this continues as the system evolves, essentially tracing collisionless orbits in the stellar potential (the $v_\infty$ distribution does not evolve substantially from 100~minutes to 7~hours).  The highest $v_\infty$ occur in the tip of the tidal tail, where lower $v_\infty$ come from slightly more-bound material closer to the surviving planetary core. We note a lower truncation of the velocity distribution at the planetary escape velocity of $v_\infty \gtrsim \sqrt{2 G m_{\rm planet}/R_{\rm planet}} \sim 11$~km~s$^{-1}$. Slower moving material remains bound to the surviving planetary core. 

The peak of the velocity distribution, clearly visible in the histogram of the right-hand panel of Figure \ref{fig:vinf}, has a magnitude similar to that predicted by $v_\infty \sim \sqrt{2 \Delta E}$, shown with a vertical dashed line. It is slightly higher due to the nonlinear distortion of the planet at periapse. Because the planet is already stretched by the time it reaches periapse, the tide it experiences is slightly stronger than that estimated based on its original radius. 

Finally, the right panel of Figure \ref{fig:vinf} compares the velocity distribution of material from the outer 10\% of the planetary radius as compared to all unbound debris. We observe that the fastest moving tail of material is largely made up of the outermost material in the planet. This effect highlights what may be important for the chemical--kinematic output in the tidal disruption of planets: the fastest debris comes from the outer planetary layers. In the context of chemically-differentiated rocky planets, this may mean that a large fraction of the fastest moving interstellar debris comes from outer crust or mantle layers. 

Fully disruptive tidal encounters, those without a surviving planetary core, occur in deeper periapse passages with $r_{\rm peri} \lesssim r_{\rm t}$. In these cases, a broader distribution of $v_\infty$ is expected, as is an extension down to $v_\infty\sim 0$ \citep{2013ApJ...767...25G}.  We note that the particular distribution of mass  in binding energy, and whether a core survives a particular encounter, is sensitive to the interior structure of the planet, not just its average density. The effects of planetary structure and thresholds for disruption have been previously explored in numerical simulations by \citet{2005Icar..175..248F,2011ApJ...732...74G,2013ApJ...762...37L,Per1}. In general, partial tidal disruptions that leave surviving cores occur out to about twice the periapse distance that leads to complete disruptions of the planet. If chemical stratification accompanies a layered planetary interior structure, this will impact the composition of debris as a function of impact parameter \citep{Bro23}. Given these broader considerations,  we defer a numerical parameter study to later work, which could also consider a more detailed treatment of the rocky-planet equation of state.   However, Section 4.2 will consider the impact of periapse distance on the asymptotic velocity of debris.

\section{Interstellar Population of Unbound Debris}

Next, we considers the implications of many tidal disruption events by dwarf stars in creating a population of interstellar objects. 

\subsection{The IM1 Interstellar Object}
On 8 January 2014, US government satellite sensors detected three atmospheric flares separated by a tenth of a second from each other, $\sim$84~km north of Manus Island. Analysis of the trajectory suggested an interstellar origin of the meteor CNEOS-2014-01-08 (hereafter IM1), with an arrival speed relative to the Local Standard of Rest of the Milky-Way galaxy, $\sim 60~{\rm km~s^{-1}}$, higher than that of 95\% of the stars in the Sun’s vicinity~\citep{SL22a} 

In 2022 the US Space Command issued a formal letter to NASA certifying a 99.999\% likelihood that the object was interstellar in origin.~\footnote{\href{https://lweb.cfa.harvard.edu/~loeb/DoD.pdf}{https://lweb.cfa.harvard.edu/\textasciitilde loeb/DoD.pdf}} Along with this letter, the US Government released the fireball lightcurve as measured by satellites.~\footnote{\href{https://lweb.cfa.harvard.edu/~loeb/lightcurve.pdf}{https://lweb.cfa.harvard.edu/\textasciitilde loeb/lightcurve.pdf}} IM1 broke apart at an unusually low altitude of $\sim$17 km, corresponding to a ram pressure of $\sim 200$ MPa.
This suggested that the object was substantially stronger than any of the other 272 bolides in the CNEOS catalog - which all disintegrated at lower ram pressures, including the $\sim$5\% minority of iron meteorites from the solar system~\citep{SL22b}. Calculations of the fireball light energy suggest that about 500 kg of material was ablated by the fireball and converted into spherules with a small efficiency~\citep{TR22}. The fireball path was localized to a $\sim$1 km-wide strip based on the delay in arrival time of the direct and reflected sound waves to a seismometer located on Manus Island~\citep{2023arXiv230307357S}. 

A towed-magnetic-sled survey during the period 14-28 June, 2023, over the seafloor about 80-90 km north of Manus Island found about 850 spherules of diameter 0.05-1.3 millimeters~\citep{Loeb2023,Loeb24a,Loeb24c,Loeb24b}. An excess of spherules was found along the expected meteor path. Mass spectrometry of the spherules along IM1's path revealed a distinct extra-solar abundance pattern for $\sim (2-10)$\% of them, while background spherules showed abundances consistent with a solar system origin. The unique spherules showed an excess of Be, La and U, by up to three orders of magnitude relative to the solar system standard of CI chondrites. These ``BeLaU"-type
spherules, also have very low refractory siderophile elements such as Re. Volatile elements, such as Mn, Zn, Pb, are depleted
as expected from evaporation losses during a meteor's airburst. The ``BeLaU"-type abundance pattern, never reported before for solar system materials, points towards association with IM1, and supports IM1's interstellar origin independently of the high velocity and unusual material strength implied from the CNEOS data.

\subsection{Velocities}

\begin{figure}
    \centering
    \includegraphics[width=\columnwidth]{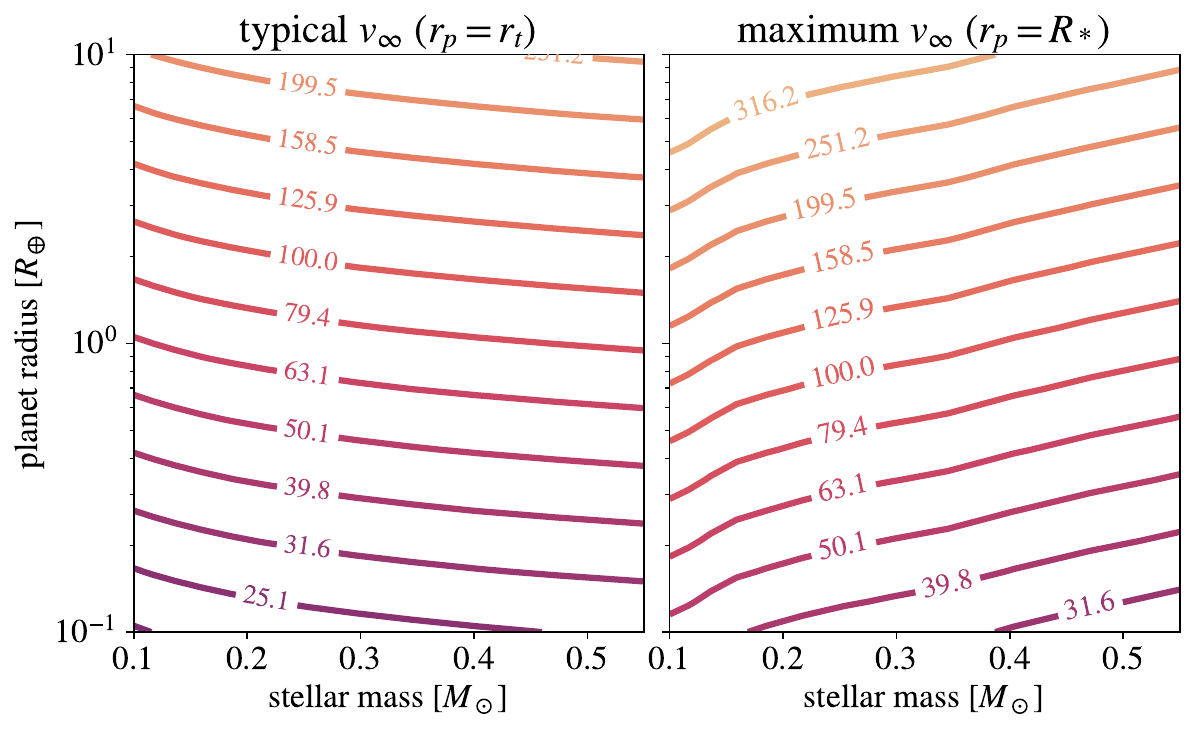}
    \caption{Asymptotic debris velocities estimated for passages at the tidal radius (left panel) and at the stellar photosphere radius (right panel) -- which represents a mininum periapse distance and maximum debris velocity for a given star-object combination. We show the effects of a range of stellar masses and planetary radii, assuming a planetary density of 5~g~cm$^{-3}$.  }
    \label{fig:popvel}
\end{figure}

Remarkably, the typical asymptotic speed estimated from equation (\ref{eq:60}) or simulated in Section \ref{sec:sim} matches the inferred interstellar speed of IM1 relative to the Local Standard of Rest -- which otherwise represents the 5\% outlier tail in the  local velocity ellipsoid of stars~\citep{Sun23}.
We conclude that the excess energy imparted to the tidal stream of rocks around common M-dwarfs with $M_\star\sim 0.1M_\odot$ naturally accounts for the inferred IM1 speed of $\sim 60~{\rm km~s^{-1}}$ relative to the Local Standard of Rest~\citep{SL22a}. Guided by this comparison, we consider the range of possible velocities from tidal encounters between main sequence stars and rocky objects and the resulting population of interstellar objects. 

Figure \ref{fig:popvel} considers the range of possible asymptotic velocities from tidal encounters. Given hypothetical planets, moons and planetesimals of density 5~g~cm$^{-3}$, we show the predicted velocity from equation (\ref{eq:60}) given a periapse distance at the tidal radius (left panel) or at the stellar photosphere (right panel). Encounters at the tidal radius represent a typical disruptive encounter because in most dynamical processes encounters with smaller periapse distances are expected to be more rare.
Encounters with $r_{\rm peri}\sim R_*$ represent the maximum possible velocity imparted; smaller periapse distances lead to the star engulfing the planet. 

In general, larger planets generate tidal debris that is faster moving. Planetesimals on the order of Ceres' size ($\sim 0.1 R_\oplus$) lead to velocities in the range of 20-40~km~s$^{-1}$. By contrast, a large planet of several Earth radii could generate debris in excess of 100~km~s$^{-1}$. However, a robust conclusion is that across the range of possible rocky object or stellar properties, asymptotic speeds in the range of tens to a hundred km~s$^{-1}$ are typical  \citep{Per1,Per2}. This is suggestive that the inferred IM1 speed is not coincidental. We note that rare, very close encounters WDs could generate tidal debris that is faster moving  -- up to a few thousand km~s$^{-1}$ in the most extreme encounters with massive WDs \citep{2018ApJ...861...35R,Per2}.

\subsection{Population of Interstellar Debris}

In the limit of high eccenticity encounters, where $\Delta E \gg |E_{\rm orb}|$, we have argued that about half of the tidal debris would be expelled to interstellar space. If it is not already molten, it may be melted by either the tidal strain or the stellar radiance -- particularly if a planet were to suffer many close passages prior to disruption.  By comparison to MIST models of main sequence stars, the equilibrium temperature at $r_{\rm peri} \sim r_{\rm t}$ is 
\begin{equation}\label{eq:Teq}
    T_{\rm eq, peri} \sim 1650~{\rm K} \left( M_* \over 0.1 M_\odot \right)^{1/2},
\end{equation}
where the scaling is representative for stars $\lesssim 0.5 M_\odot$. 

However, as the material expands in the tidal tails, it cools. Along the way, it solidifies, forms clumps, and becomes a collection of solid objects rather than a fluid distribution.  Modeling the physics of this process is complex but certainly merits deeper consideration than we are able to provide. Relevant processes might include the self-gravity of debris, the thermodynamics of its cooling liquid-to-solid and state-change, and perhaps even the surface tension of coalescing fragments within the expanding background of the debris tail \citep[e.g.][]{2018ApJ...861...35R,Zhang20,2022ApJ...931L...6B,2023MNRAS.522.5500C,2023MNRAS.526.2323F}. 

One empirical hint at the fragments that might result is IM1 itself -- if we presume its origin from this process -- which has a mass of $\sim 500$~kg. Other empirical evidence comes from the observed fragment size distribution of bound debris  around WDs~\citep{Van15a,Ver21,Ver24}, which might be expected to be subject to a similar cooling and fragmentation process as the unbound tail. However, the bound disk is subject to a phase of collisional fragmentation during its circularization which likely doesn't apply to the unbound debris.

We can estimate the interstellar population of IM1-like objects based on the  density of host stars and the mass of expelled debris per star, $M_{\rm rock}$. We assume that $M_{\rm rock}$ is comparable to the retained mass in rocks around M-dwarfs, which implies that $M_{\rm rock}$ is half the total mass of disrupted planets. We then assume that this mass is divided into IM1-mass rocks with $\sim 500$~kg, which lets us convert the  local number density of M-dwarfs~\citep{Winters21} $n_{\rm MD}\approx 0.1~{\rm pc^{-3}}$, to an estimated number density of  IM1-like rocks per Earth mass of expelled rocky material by each M-dwarf
\begin{equation}
    n_{\rm IM1}\sim 1.5\times 10^5~{\rm au}^{-3} \left( {M_{\rm rock} \over M_\oplus}\right)
\end{equation} 
The mass of disrupted planets, and therefore $M_{\rm rock}$, accrued over the lifetime of an M-dwarf are unknown, but we can be guided to estimates based on the census of rocky planets that are gravitationally-bound to nearby M-dwarfs. Typical planetary systems appear to contain more than a few Earth masses of rocky planets. If disruptions are common, as the evidence from polluted WDs suggests, a similar mass of planets might be tidally disrupted, with their material contributing to interstellar debris.  For example, the total mass in reported rocky planets around the nearest M-dwarf, Proxima Centauri ($M_\star=0.12M_\odot$), is $(1.3+7+0.3)M_\oplus \sim 8.6M_\oplus$~\citep{Anglada16,Damasso20,Faria22}, and the seven reported planets around TRAPPIST-1 ($M_\star=0.09M_\odot$) total a mass of, $(1.4+1.3+0.4+0.7+1+1.3+0.3)M_\oplus=6.4M_\oplus$~\citep{Gillon17}. Other nearby M dwarfs are known or suspected to host planetary systems~\citep{Ribas18,Bonfils18,Jeffers20,Diaz19,Tuomi19,Lillo20}.
Dynamical instability was recently explored for the multi-planet system of the low-mass ($0.36M_\odot$) star GJ 357 which contains $\gtrsim 11M_\oplus$ in planets~\citep{Kane23}. 

Given these considerations, the implied collision rate of IM1-like meteors on Earth is
\begin{equation}
\Gamma_{\rm IM1}= n_{\rm IM1}\times (\pi R_\oplus^2) \times v_\infty \sim 0.1\left({M_{\rm rock}\over 10M_\oplus}\right)~{\rm yr}^{-1},
\label{eq:rate}
\end{equation}
where we have scaled our estimate to 10 Earth masses of expelled rock by each M dwarf based on the discussion above (implying a disrupted mass of planets of 2$M_{\rm rock} \sim 20 M_\oplus$). 
The resulting rate of interstellar meteors with a mass of $\sim 500$kg is comparable to the IM1 detection rate of once per decade in the CNEOS catalog.~\footnote{\href{https://cneos.jpl.nasa.gov/fireballs/}{https://cneos.jpl.nasa.gov/fireballs/}}

Because the true size distribution is unknown, we might alternatively consider scale-free fragmentation through collisions in the tidal stream, which would lead to equal amount of mass per logarithmic interval in fragment mass. As a specific example, if the debris spans ten orders of magnitude in fragment mass the estimated mass of fragments similar to IM1 would be lower by a factor of ten, but lower-mass fragments would be much more numerous.

\subsection{Compositional and Kinematic Differentiation}

The chance to directly examine an interstellar sample in IM1 has lead to insights into its unique composition~\citep{Loeb2023,Loeb24a,Loeb24c,Loeb24b}, which appears most similar to differentiated crustal planetary material. It has enhanced abundances of Be, La and U relative to the standard composition of CI chondrites, suggested by the ``BeLaU"-type composition of the spherules collected near IM1’s path~\citep{Loeb2023,Loeb24a,Loeb24c,Loeb24b}. 

This abundance pattern could be a remnant of the properties of the source planet, or something that occurred during periapse passage. In our own solar system, many rocky objects are differentiated -- from Ceres to the terrestrial planets and moons. This differentiation is a result of the initial phases during which the objects were hot enough to support magma oceans before they cooled \citep[e.g.][]{2015GMS...212...83N}. 

Intriguingly, our simulation shows that the outermost layers of a planet are stripped and expelled with the highest velocities. Though the process of tidal disruption is violent, it does not mix the ejecta. Thus, chemically differentiated planets being tidally disrupted lead to kinematically differentiated debris. 

In principle, the rate of elemental differentiation could be accelerated during many periapse passages on highly eccentric orbits, leading to a more extreme differentiation than found on the early Earth, the Moon and Mars. The ``reheating" of planets at a few times the stellar radius of M-dwarfs could result in a temperature above the melting temperature of rock $\gtrsim 1.5\times 10^3$K, as indicated by equation (\ref{eq:Teq}). 

It is possible that the high material strength implied for IM1~\citep{SL22a} could be the result of ``BeLaU"-type composition and multiple heating episodes during many periapse passages before their ultimate tidal disruption near their parent M-dwarfs. It is also possible that this represents interplanetary variance and the evolutionary history of a planetary system other than our own. More detailed models of the thermodynamics of a periapse passage, might be able to constrain these scenarios further.

\section{Conclusions}

We have shown that rocky bodies can be expelled from their host stars when planets or asteroids are scattered into eccentric orbits and tidally disrupted by their host stars. In particular, the most common stars, M-dwarfs, are high enough density to tidally disrupt typical rocky planets. 

Rocky planets might develop high eccentricity orbits as a result of a secular torques from an outer (possibly giant) planets or stellar binary companions. Empirically, there is evidence that this process occurs frequently from the pollution observed on WDs. 

The excess energy imparted to the tidal stream from the disruption of a rocky planet with the size of the Earth near common M-dwarfs of mass $M_\star\sim 0.1M_\odot$, leads to a typical speed of $\sim 60~{\rm km~s^{-1}}$ relative to the Local Standard of Rest (equation \ref{eq:60} and Figures \ref{fig:vinf} and \ref{fig:popvel}), quite similar to that inferred from the properties of the candidate interstellar meteor IM1.  For $\sim 10M_\oplus$ rocky reservoirs around M-dwarfs, the tidal disruption of rocky material accounts for a collision rate of 500kg-mass meteors of once per decade (Eq.~\ref{eq:rate}), consistent with that of IM1. 

Our models show that any differentiation in a disrupted planet is preserved in the tidal debris tails, with the outermost layers ejected fastest. This may account for the ``BeLaU"-type composition of unique spherules along IM1’s path~\citep{Loeb2023,Loeb24a,Loeb24c,Loeb24b}. Melting of the rock during tight periapse passages at a few times the stellar radius, may also naturally result in additional elemental differentiation and enhanced abundances of Be, La and U in the crust, which is eventually disrupted and expelled.  

Our model has not considered the detailed equation of state or material strength of rocky planets. Nor have we made detailed arguments about the size distribution of unbound debris. Both of these physical processes may be important in shaping the interstellar population from unbound debris of planetary disruption and merit further consideration.

\begin{acknowledgements}
We acknowledge support from the Galileo Project at Harvard University. MM gratefully acknowledges support from the Clay Postdoctoral Fellowship of the Smithsonian Astrophysical Observatory. This work used stampede2 at the Texas Advanced Computing Center through allocation PHY230031 from the Advanced Cyberinfrastructure Coordination Ecosystem: Services \& Support (ACCESS) program, which is supported by National Science Foundation grants \#2138259, \#2138286, \#2138307, \#2137603, and \#2138296.
\end{acknowledgements}

Software to reproduce the analysis presented in this paper is archived online at \url{https://github.com/morganemacleod/PlanetTDEMaterials}. 

\bibliographystyle{aa}
%\bibliography{refs.bib}

\end{document}